\documentclass[aps,twocolumn,floatfix,superscriptaddress]{revtex4}
\usepackage{amsmath}
\usepackage{graphicx}
\usepackage{bm}
\begin{document}
\title{Parity and Predictability of Competitions}
\author{E.~Ben-Naim}
\email{ebn@lanl.gov}
\affiliation{Theoretical Division and Center for Nonlinear Studies,
Los Alamos National Laboratory, Los Alamos, New Mexico 87545}
\author{F.~Vazquez}
\email{fvazquez@buphy.bu.edu}
\affiliation{Theoretical Division and Center for Nonlinear Studies,
Los Alamos National Laboratory, Los Alamos, New Mexico 87545}
\affiliation{Department of Physics, Boston University, Boston,
Massachusetts 02215}
\author{S.~Redner}
\email{redner@bu.edu}
\affiliation{Theoretical Division and Center for Nonlinear Studies,
Los Alamos National Laboratory, Los Alamos, New Mexico 87545}
\affiliation{Department of Physics, Boston University, Boston,
Massachusetts 02215}
\begin{abstract}
We present an extensive statistical analysis of the results of all
sports competitions in five major sports leagues in England and the
United States. We characterize the parity among teams by the variance
in the winning fraction from season-end standings data and quantify
the predictability of games by the frequency of upsets from game
results data. We introduce a novel mathematical model in which the
underdog team wins with a fixed upset probability. This model
quantitatively relates the parity among teams with the predictability
of the games, and it can be used to estimate the upset frequency from
standings data.
\end{abstract}

\maketitle

What is the most competitive sports league?  We answer this question
via an extensive statistical survey of game results in five major
sports. Previous studies have separately characterized {\it parity}
(Fort 1995) and {\it predictability} (Stern 1997, Wesson 2002, Lundh
2006) of sports competitions. In this investigation, we relate parity
with predictability using a novel theoretical model in which the
underdog wins with a fixed upset probability.  Our results provide
further evidence that the likelihood of upsets is a useful measure of
competitiveness in a given sport (Wesson 2002, Lundh 2006).  This
characterization complements the myriad of available statistics on the
outcomes of sports events (Albert 2005, Stern 1991, Gembris 2002).

We studied the results of nearly all regular season competitions in 5
major professional sports leagues in England and the United States
(table I): the premier soccer league of the English Football
Association (FA), Major League Baseball (MLB), the National Hockey
League (NHL), the National Basketball Association (NBA), and the
National Football League (NFL). NFL data includes the short-lived
AFL. Incomplete seasons, such as the quickly abandoned 1939 FA season,
and nineteenth-century results for the National League in baseball
were not included.  In total, we analyzed more than 300,000 games in
over a century (data source: http://www.shrpsports.com/,
http://www.the-english-football-archive.com/).

\section{Quantifying Parity}

The winning fraction, the ratio of wins to total games, quantifies
team strength.  Thus, the distribution of winning fraction measures
the parity between teams in a league.  We computed $F(x)$, the
fraction of teams with a winning fraction of $x$ or lower at the end
of the season, as well as $\sigma=\sqrt{\langle x^2\rangle -\langle
x\rangle^2}$, the standard deviation in winning fraction. Here
$\langle\cdot\rangle$ denotes the average over all teams and all years
using season-end standings. In our definition, $\sigma$ gives a
quantitative measure for parity in a league (Fort 1995, Gould
1996). For example, in baseball, where the winning fraction $x$
typically falls between $0.400$ and $0.600$, the variance is
$\sigma=0.084$.  As shown in figures 1 and 2a, the winning fraction
distribution clearly distinguishes the five leagues. It is narrowest
for baseball and widest for football.

\begin{figure}[t]
\vspace*{0.01cm}
\includegraphics*[width=0.37\textwidth]{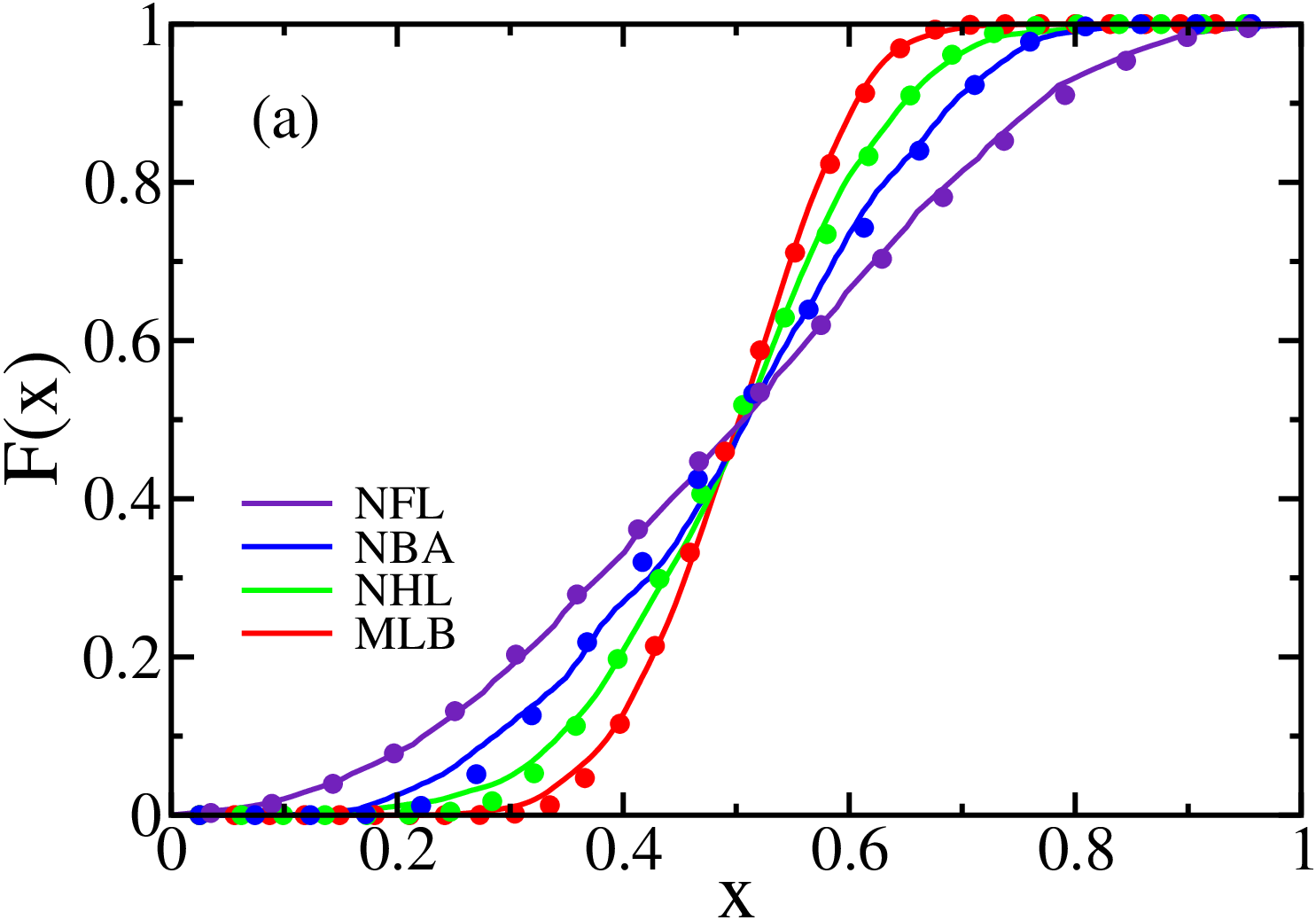}
\caption{Winning fraction distribution (curves) and the best-fit
distributions from simulations of our model (circles). For clarity,
FA, that lies between MLB and NHL, is not displayed.}
\end{figure}

Do these results imply that MLB games are the most competitive and NFL
games the least? Not necessarily! The length of the season is a
significant factor in the variability in the winning fraction.  In a
scenario where the outcome of a game is random, i.e., either team can
win with equal probability, the total number of wins performs a simple
random walk, and the standard deviation $\sigma$ is inversely
proportional to the square root of the number of games
played. Generally, the shorter the season, the larger $\sigma$. Thus,
the small number of games is partially responsible for the large
variability observed in the NFL.

\begin{table}[t]
\begin{tabular}{|l|c|c|c|c|c|c|}
\hline
league&years&games&$\langle{\rm games}\rangle$&$\sigma$&${\bm q}$&$q_{\rm model}$\\
\hline
FA&1888-2005&43350&39.7&0.102&{\bf 0.452}&0.459\\
MLB&1901-2005&163720&155.5&0.084&{\bf 0.441}&0.413\\
NHL&1917-2004&39563&70.8&0.120&{\bf 0.414}&0.383\\
NBA&1946-2005&43254&79.1&0.150&{\bf 0.365}&0.316\\
NFL&1922-2004&11770&14.0&0.210&{\bf 0.364}&0.309\\
\hline
\end{tabular}
\caption{Summary of the sports statistics data. Listed are the time
  periods, total number of games, average number of games played by a
  team in a season ($\langle{\rm games}\rangle$), variance in the
  win-percentage distribution ($\sigma$), measured frequency of upsets
  (${\bm q}$), and upset probability obtained using the theoretical
  model ($q_{\rm model}$). The fraction of ties in soccer, hockey, and
  football is $0.246$, $0.144$, and $0.016$, respectively.}
\label{table}
\end{table}

\section{Quantifying Predictability}

To account for the varying season length and reveal the true nature of
the sport, we set up artificial sports leagues where teams, paired at
random, play a fixed number of games. In this simulation model, the
team with the better record is considered as the favorite and the team
with the worse record is considered as the underdog.  The outcome of a
game depends on the relative team strengths: with ``upset
probability'' $q<1/2$, the underdog wins, but otherwise, the favorite
wins. If the two teams have the same fraction of wins, one is randomly
selected as the winner.

We note that a similar methodology was utilized by Wesson who focused
on the upset likelihood as a function of the final point spread in
soccer (Wesson 2002). Also, an equivalent definition of the upset
frequency was very recently employed by Lundh to characterize how
competitive tournaments are in a variety of team sports (Lundh 2006).

Our analysis of the nonlinear master equations that describe the
evolution of the distribution of team win/loss records shows that
$\sigma$ decreases both as the season length increases and as games
become more competitive, i.e., as $q$ increases.  This theory is
described in the appendix and more generally in Ben-Naim et al.
2006. The basic quantity to characterize team win/loss records is
$F(x)$, the fraction of teams that have a winning fraction that is
less than or equal to $x$.  In a hypothetical season with an
infinite number of games, the winning fraction distribution is
uniform
\begin{equation}
\label{phi} F(x)=
\begin{cases}
0& 0<x<q\\
{\displaystyle \frac{x-q}{1-2q}}&q<x<1-q\\
1& 1-q<x.
\end{cases}
\end{equation}
From the definition of the upset probability, the lowest winning fraction
must equal $q$, while the largest winning fraction must be $1-q$.

By straightforward calculation from $F(x)$, the standard
deviation $\sigma$ is a linear function of the upset probability
\begin{equation}
\label{sigma}
\sigma=\frac{1/2-q}{\sqrt{3}}.
\end{equation}
Thus, the larger the probability that the stronger team wins, the
greater the disparity between teams.  Perfect parity is achieved when
$q=1/2$, where the outcome of a game is completely random.  However,
for a finite and realistic number of games per season, such as those
that occur in sports leagues, we find that the variance is larger than
the infinite game limit given in Eq.~(\ref{sigma}). As a function of
the number of games, the variance decreases monotonically, and it
ultimately reaches the limiting value (\ref{sigma}).

We run numerical simulations of these artificial sports leagues by
simply following the rules of our theoretical model. In a simulated
game, the records of each team are updated according to the following
rule: if the two teams have a different fraction of wins, the favorite
wins with probability $1-q$ and the underdog wins with probability
$q$. If the two teams are equal in strength, the winner is chosen at
random. Using the simulations, we determined the value of $q_{\rm
model}$ that gives the best match between the distribution $F(x)$ from
the simulations to the actual sports statistics (figure 1). Generally,
there is good agreement between the simulations results and the data,
as quantified by $q_{\rm model}$ (table I).

\begin{figure}[t]
\vspace*{0.01cm}
\includegraphics*[width=0.37\textwidth]{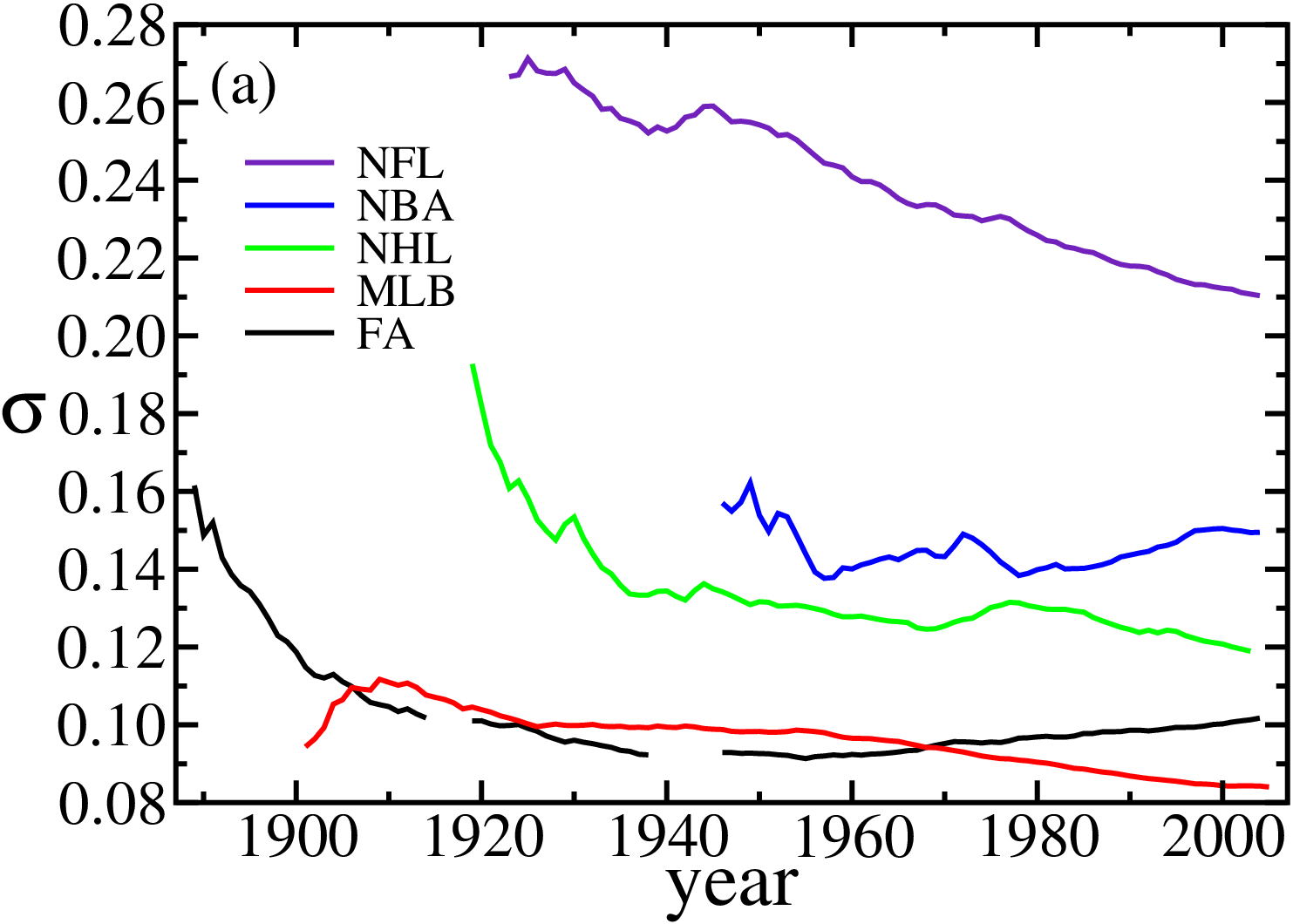}
\includegraphics*[width=0.37\textwidth]{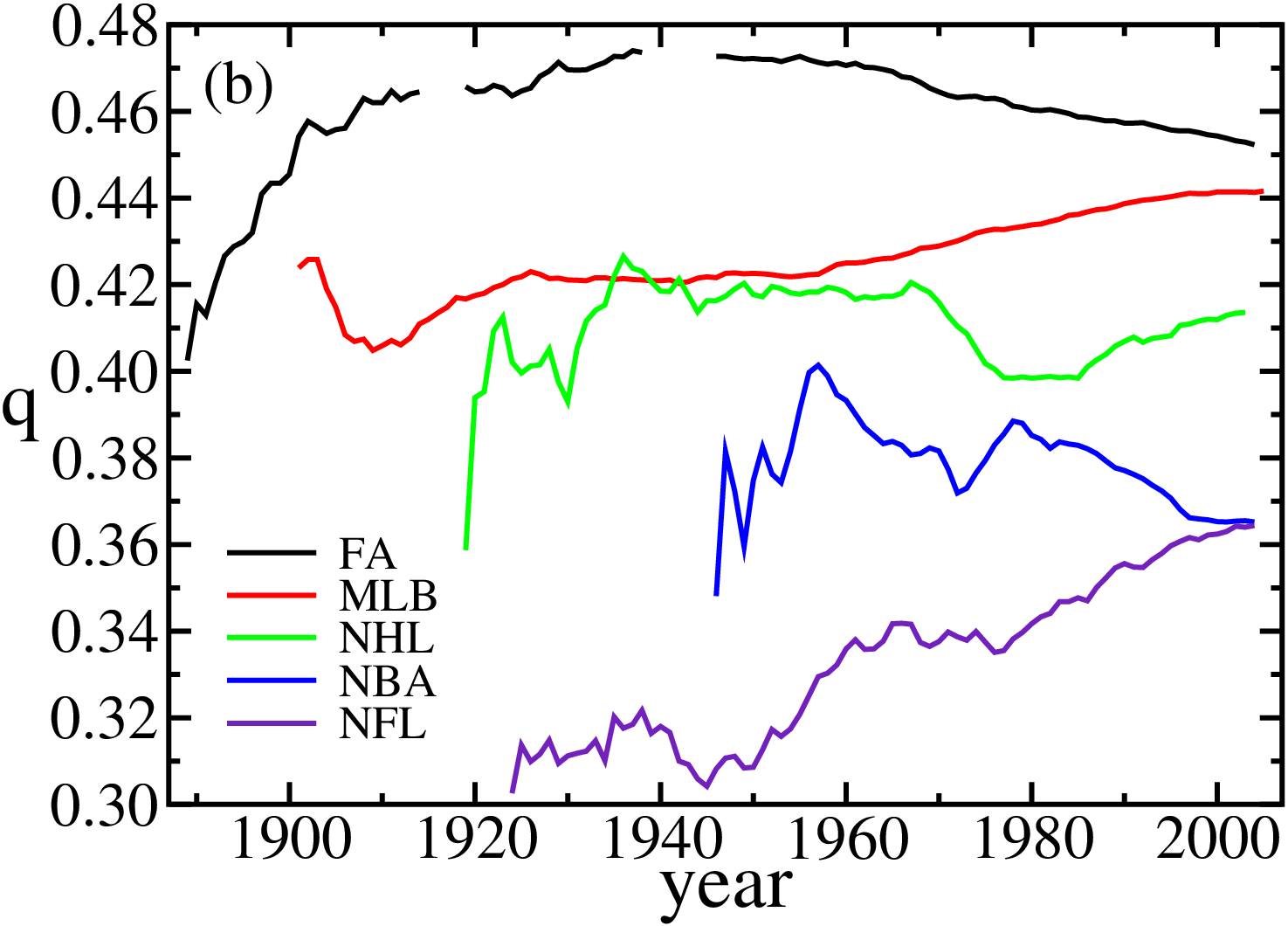}
\caption{(a) The cumulative variance in the winning fraction distribution (for
all seasons up to a given year) versus time.  (b) The cumulative
frequency of upsets $q$, measured directly from game results, versus
time.}
\end{figure}

To characterize the predictability of games directly from the game
results data, we followed the chronologically-ordered results of all
games and reconstructed the league standings at any given day. We then
measured the upset frequency $q$ by counting the fraction of times
that the team with the worse record on the game date actually won
(table I). Games between teams with no record (start of a season) or
teams with equal records were disregarded. Game location was ignored
and so was the margin of victory. In soccer, hockey, and football,
ties were counted as 1/2 of a victory for the underdog and 1/2 of a
victory for the favorite. We verified that this definition did not
have a significant affect on the results. The upset probability
changes by at most $0.02$ (and typically, much less) if ties are
ignored altogether. We note that to generalize our model to situations
with ties, it is straightforward to add a second parameter, the
probability of a tie, into the model definition.

Our main result is that soccer and baseball are the most competitive
sports with $q=0.452$ and $q=0.441$, respectively, while basketball
and football, with nearly identical $q=0.365$ and $q=0.364$, are the
least (Stern 1997, Stern 1998).

There is also good agreement between the upset probability $q_{\rm
model}$, obtained by fitting the winning fraction distribution from
numerical simulations of our model to the data as in figure 1, and the
measured upset frequency (table I). We notice however a systematic
bias in the estimated upset frequencies: the discrepancy between $q$
and $q_{\rm model}$ grows as the games become less
competitive. Consistent with our theory, the variance $\sigma$ mirrors
the bias, $1/2-q$ (figures 2a and 2b).  Tracking the evolution of
either $q$ or $\sigma$ leads to the same conclusions: (1) MLB games
have been steadily becoming more competitive (Gould 1996), (2) NFL has
dramatically improved the competitiveness of its games over the past
40 years, and (3) over the past 60 years, FA displays an opposite
trend with the games becoming less competitive.

\section{All-time team records}

In our theory, both the season length and the upset probability
affect the broadness of the win fraction distribution. However, in a
hypothetical season with an infinite number of games, the
distribution is governed by the upset probability alone. In this
case, the bias $1/2-q$ and the variance $\sigma$ are equivalent
measures of the competitiveness, as indicated by (\ref{sigma}).

\begin{figure}[h]
 \vspace*{0.cm}
\includegraphics*[width=0.42\textwidth]{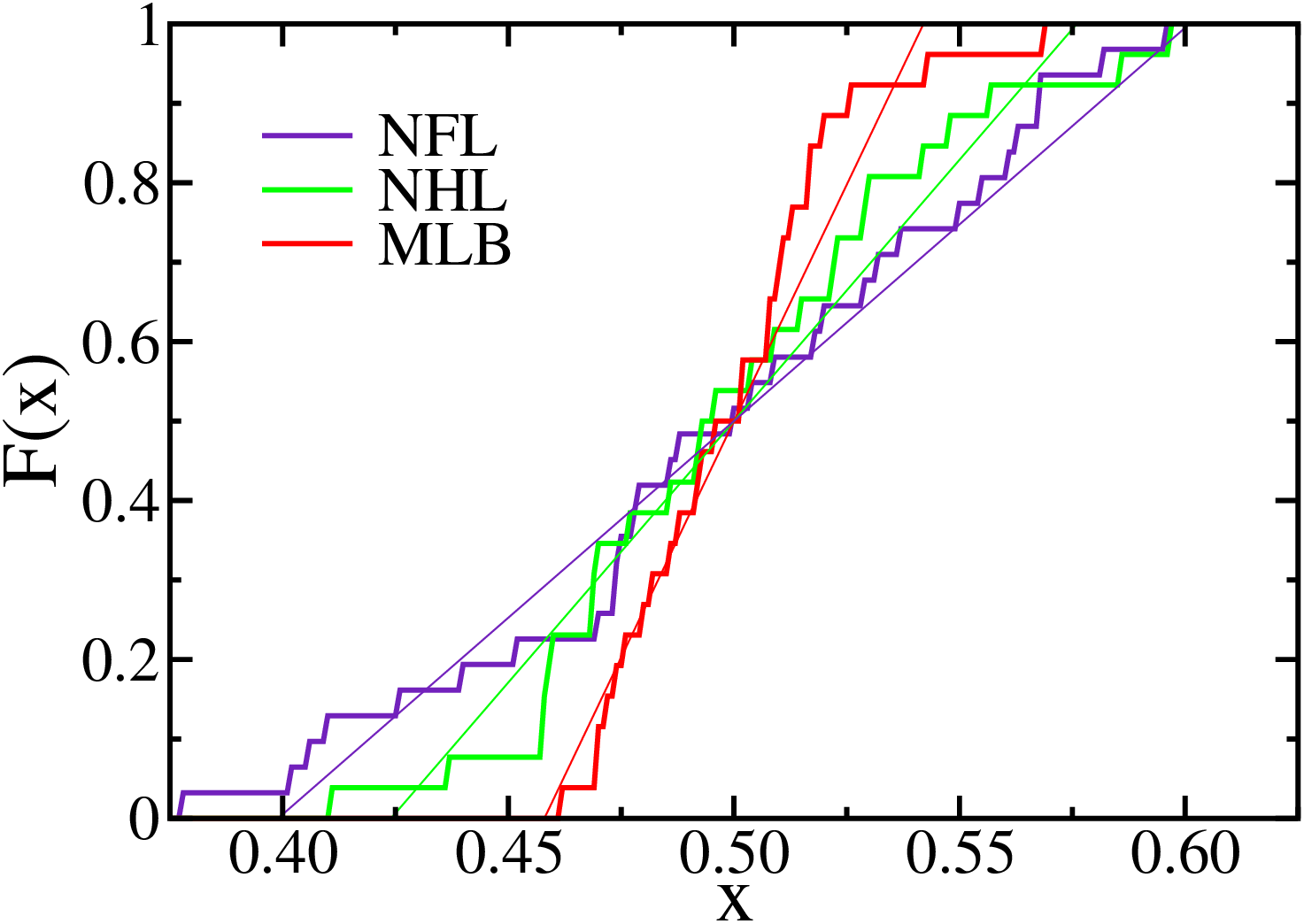}
\caption{The all-time cumulative win-fraction distribution for active
teams with 10 season minimum. For clarity, FA and NBA data is not
displayed.  The theoretical curves for an infinite season using
$q_{\rm all}$ obtained by substituting $\sigma_{\rm all}$ into
(\ref{sigma}) are shown for reference (table II).} \label{fig-tot}
\end{figure}

\begin{table}[h]
\begin{tabular}{|l|c|c|c|c|c|c|}
\hline
League&Teams &$\langle$Games$\rangle$ &$\sigma_{\rm all}$&$q_{\rm all}$&$x_{\rm max}$&$x_{\rm min}$\\
\hline
FA&37&2154&0.035&0.439&$0.582$(LPL)&$0.406$(CP)\\
MLB&26&13100&0.024&0.458&$0.567$(NYY)&$0.459$(SDP)\\
NHL&26&2850&0.044&0.424&$0.589$(MC)&$0.403$(TBL)\\
NBA&27& 3060&0.057&0.401&$0.616$(LAL)&$0.358$(LAC)\\
NFL&31& 720 &0.057&0.401&$0.600$(MD) &$0.358$(TBB)\\
\hline
\end{tabular}
\caption{Summary of the sports statistics data presented in Figure
  \ref{fig-tot}. The average number of games played by teams since
  their inception is denoted by $\langle$Games$\rangle$. The quantity
  $\sigma_{\rm all}$ is the variance in the all-time winning
  percentage of the roughly 30 sports clubs. The maximal and minimal
  fraction of wins for individual teams are indicated by $x_{\rm max}$
  and $x_{\rm min}$, respectively.  The team acronyms are: (LPL)
  Liverpool, (CP) Crystal Palace, (NYY) New York Yankees, (SDP) San
  Diego Padres, (MC) Montreal Canadiens, (TBL) Tampa Bay Lightning,
  (LAL) Los Angeles Lakers, (LAC) Los Angeles Clippers, (MD) Miami
  Dolphins, (TBB) Tampa Bay Buccaneers.}
\label{table-tot}
\end{table}

The all-time records of teams provide the longest possible win-loss
record. This comparison is of course limited by the small number of
teams, that varies between 26 and 37 (we ignored defunct franchises
and franchises participating in less than 10 seasons), and the
significant variations in the total number of games played by the
teams. Interestingly, $F(x)$ obtained from the all-time win-loss
records is reasonably close to the uniform distribution suggested by
the theory (Fig.~\ref{fig-tot} and Table \ref{table-tot}). The slope
of the line in figure \ref{fig-tot} was obtained using the theory: the
upset probability $q_{\rm all}$ was estimated from the observed
variance $\sigma_{\rm all}$ using Eq.~(\ref{sigma}). This provides
additional support for the theoretical model.

Overall, the win fraction distribution for the team all-time winning
record is in line with the rest of our findings: soccer and baseball
are the most competitive sports while basketball and football are the
least. We note that the win fraction distribution is extremely narrow,
and the closest to a straight line, for baseball because of the huge
number of games. Even though the total number of games in basketball
is four times that of football, the two distributions have comparable
widths. The fact that similar trends for the upset frequency emerge
from game records as do from all-time team records indicate that the
relative strengths of clubs have not changed considerably over the
past century.

\section{Discussion}

In summary, we propose a single quantity, $q$, the frequency of
upsets, as an index for quantifying the predictability, and hence the
competitiveness of sports leagues.  This quantity complements existing
methods addressing varying length seasons and in particular,
competitive balance that is based on standard deviations in winning
percentages (Fort 2003).  We demonstrated the utility of this measure
via a comparative analysis that shows that soccer and baseball are the
most competitive sports. Trends in this measure may reflect the
gradual evolution of the teams in response to competitive pressure
(Gould 1996, Lieberman 2005), as well as changes in game strategy or
rules (Hofbauer 1998). What plays the role of fitness in this context
is in open question.

In our definition of the upset frequency we ignored issues associated
with unbalanced schedules, unestablished records, and variations in
the team strengths. For example, we count a game in which a 49-50 team
beats a 50-49 team as an upset. To assess the importance of this
effect we ignored all games between teams separated by less than
$0.05$ in win-percentage.  We find that the upset frequency changes by
less than $0.005$ on average for the five sports. Also, one may argue
that team records in the beginning of the season are not well
established and that there are large variations in schedule
strength. To quantify this effect, we ignored the first half of the
season. Remarkably, this changes the upset frequency by less than
$0.007$ on average. We conclude that issues associated with strength
of schedule and unbalanced schedules have negligible influence on the
upset frequency.

It is worth mentioning that our model does not account for several
important aspects of real sports competitions.  Among the plethora of
such issues, we list a few prominent examples: (i) Game location. Home
and away games are not incorporated into our model, but game location
does affect the outcome of games.  For example, during the 2005
baseball season 54\% of the total team wins occurred at home.  (ii)
Unbalanced schedule. In our fixed-game algorithm, each team plays all
other teams the same number of times.  However, some sports leagues
are partitioned into much smaller subdivisions, with teams playing a
larger fraction of their schedule against teams in their own
subgroup. This partitioning is effectively the same as reducing the
number of teams, an effect that we found has a small influence on the
distribution of win fraction. (iii) Variable upset probability.  It is
plausible that the upset probability $q$ depends on the relative
strengths of the two competing teams. It is straightforward to
generalize the model such that the upset frequency depends on the
relative strengths of the two teams and this may be especially
relevant for tournament competitions.

Despite all of these simplifying assumptions, we see the strength of
our approach in its simplicity. Our theoretical model involves a
single parameter and consequently, it enables direct and unambiguous
quantitative relation between parity and predictability.

Our model, in which the stronger team is favored to win a game,
enables us to take into account the varying season length and this
model directly relates parity, as measured by the variance $\sigma$
with predictability, as measured by the upset likelihood $q$. This
connection has practical utility as it allows one to conveniently
estimate the likelihood of upsets from the more easily-accessible
standings data. In our theory, all teams are equal at the start of the
season, but by chance, some end up strong and some weak. Our idealized
model does not include the notion of innate team strength;
nevertheless, the spontaneous emergence of disparate-strength teams
provides the crucial mechanism needed for quantitative modeling of the
complex dynamics of sports competitions.

One may speculate on the changes in competitiveness over the years.
In football there is a dramatic improvement in competitiveness
indicating that actions taken by the league including revenue sharing,
the draft, and unbalanced schedules with stronger teams playing a
tougher schedule are very effective. In baseball, arguably the most
stable sport, the gentle improvement in competitiveness may indeed
reflect natural evolutionary trends.  In soccer, the decrease in
competitiveness over the past 60 years indicate a ``rich gets richer''
scenario.

\acknowledgements We thank Micha Ben-Naim for his dedicated
assistance in data collection, Michael Nieto for suggesting
inclusion of soccer, and Harvey Rose for critical reading of the
manuscript. We acknowledge support from DOE (W-7405-ENG-36) and NSF
(DMR0227670 \& DMR0535503).

\appendix

\section{The theoretical model}

In our model, there are $N$ teams that compete against each other.
In each game there is one winner and one loser, and no ties are
possible. In each competition, the team with the larger number of
wins is considered as the favorite, and the other team as the
underdog.  The winner of each competition is determined by the
following rule: the underdog wins with upset probability $q$, and
the favorite team wins with probability $p=1-q$. If the two
competing teams have identical records, the winner is chosen
randomly.

Let $k$ be the number of wins of a team.  Then the outcome of a game
is as follows: when $k>j$
\begin{eqnarray*}
\label{rule}
(k,j)&\to& (k,j+1) \quad{\rm with\ probability~} q,\\
(k,j)&\to& (k+1,j) \quad{\rm with\ probability~} 1-q.
\end{eqnarray*}

Our theoretical analysis is based on a kinetic approach.  We set the
competition rate to $1/2$, so that the time increases by one when
every team plays one game, on average.  Then the average number of
games played by a team, $t$, plays the role of time. Also, we take the
limit of large $t$ so that fluctuations in the number of games vanish.

Let $g_k(t)$ be the fraction of teams with $k$ wins at time $t$. We
address the case where any two teams are equally likely to play
against each other. Then, the win-number distribution obeys the
master equation (Ben-Naim 2006)
\begin{equation}
\label{dis-eq-0}
\begin{split}
\frac{dg_k}{dt}&=(1-q)(g_{k-1}G_{k-1}-g_kG_k)\\
&+q(g_{k-1}H_{k-1}-g_kH_k)+\frac{1}{2}\left(g_{k-1}^2-g_k^2\right)\,.
\end{split}
\end{equation}
Here $G_k=\sum_{j=0}^{k-1} g_j$ and $H_k=\sum_{j=k+1}^\infty g_j$ are
the respective cumulative distributions of teams with less then or
more than $k$ wins.  Of course $G_k+H_{k-1}=1$.  The boundary
condition is $g_{-1}(t)=0$.  The first pair of terms describes games
where the stronger team wins, and the second pair of terms accounts
for interactions where the weaker team wins. The last pair of terms
describes games between two equal teams. The prefactor $1/2$ arises
because there are half as many ways to chose equal teams as there are
for different teams. We consider the initial condition where all teams
are equal, $g_k(0)=\delta_{k,0}$.

By summing the rate equation (\ref{dis-eq-0}), the cumulative
distribution obeys the master equation
\begin{equation}
\label{dis-eq}
\frac{dG_k}{dt}=q(G_{k-1}-G_k)+(1/2-q)\left(G^2_{k-1}-G^2_k\right).
\end{equation}
The boundary conditions are $G_{0}=0$, $G_\infty=1$, while the
initial condition for the start of each season is $G_k(0)=1$ for $k>
0$. It is simple to verify, by summing the master equations, that
the average number of wins $\langle k\rangle =\sum_k k
(G_k-G_{k-1})$, obeys $d\langle k\rangle/dt=1/2$; therefore, the
average number of wins by a team is half the number of games it
plays, $\langle k\rangle=t/2$, as it should.

\begin{figure}[t]
\includegraphics*[width=0.38\textwidth]{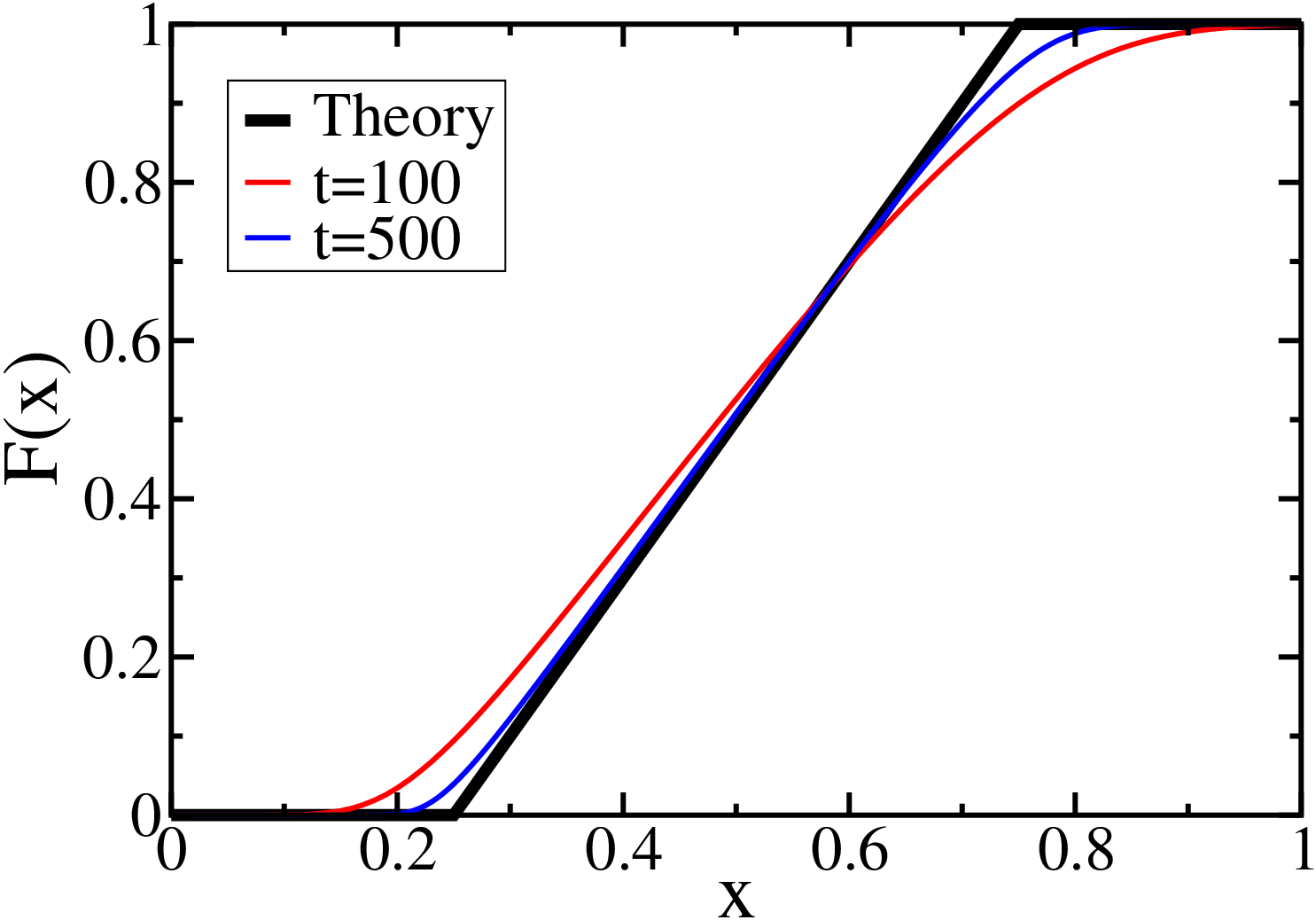}
\caption{The win-fraction distribution for $q=1/4$ at different
times $t=100$ and $t=500$.} \label{fig-phi}
\end{figure}

When the number of games is large, $t\to \infty$, we can solve the
master equation using a simple scaling analysis. Let us take the
continuum limit of the master equation by replacing differences with
derivatives, $G_{k+1}-G_k\to \partial G/\partial k$. To first order
in this ``spatial'' derivative, we obtain the nonlinear partial
differential equation
\begin{equation}
\label{cum-eq-cont} \frac{\partial G}{\partial
t}+\left[q+(1-2q)G\right] \frac{\partial G}{\partial k}=0.
\end{equation}
Since the number of wins is proportional to the number of games
played, $k\sim t$, we focus on the fraction of wins $x=k/t$. The
corresponding win-fraction distribution
\begin{equation}
\label{scaling}
G_k(t)\to F(k/t)
\end{equation}
becomes stationary in the long-time limit, $t\to\infty$.  The
boundary conditions for the win-fraction distribution is $F(0)=0$
and $F(1)=1$.

Substituting the scaled cumulative win-fraction distribution
(\ref{scaling}) into the continuum equation (\ref{cum-eq-cont}), we
find that the scaled cumulative win-fraction distribution obeys the
ordinary differential equation
\begin{equation}
\label{scaling-eq} \left[(x-q)-(1-2q)F(x)\right]\frac{dF}{dx}=0.
\end{equation}
Here the prime denotes differentiation with respect to $x$.  The
solution is either a constant $F(x)={\rm constant}$, or the linear
function $F(x)=\frac{x-q}{1-2q}$.  Using these two solutions, invoking
the boundary conditions $F(0)=0$ and $F(1)=1$, as well as continuity
of the cumulative distribution, we deduce that the winning fraction
has the form that is given in equation (\ref{phi}).  In a hypothetical
season with an infinite number of games, the win-fraction distribution
$f(x)=F'(x)$ is uniform, $f(x)=(1-2q)^{-1}$, in the range $1<x<1-q$,
while $f(x)$ vanishes outside this range.  As shown in figure 4,
numerical integration of the master equation (\ref{dis-eq}) confirms
the scaling behavior (\ref{phi}): as the number of games increases,
the win-fraction distribution approaches the limiting uniform
distribution.


\begin{thebibliography}{99}

\bibitem{abc} Albert, J., Bennett, J., and Cochran, J.J., eds.\ {\it
    Anthology of Statistics in Sports}, (SIAM, Philadelphia, 2005).

\bibitem{bvr} Ben-Naim, E., Vazquez, F., and Redner, S., ``On the
structure of competitive societies,'' Eur.\ Phys.\ Jour. B. {\bf 26}, 531
(2006).

\bibitem{fq} Fort, R. and Quirk, J., ``Cross-subsidization,
incentives, and outcomes in professional team sports leagues,''
J. Econ.\ Liter.\ {\bf 33}, 1265 (1995).

\bibitem{fm} Fort, R. and Maxcy, J., "Competitive Balance in Sports Leagues: An
Introduction," Journal of Sports Economics {\bf 4}, 154 (2003).

\bibitem{gts} Gembris, D., Taylor, J.G., and Suter, D.,
``Sports statistics - Trends and random fluctuations in athletics,''
 Nature {\bf 417}, (2002).

\bibitem{G} Gould, S.J., ``Full house: The spread of excellence from Plato
    to Darwin,'' (Harmony Books, New York, 1996).

\bibitem{hs} Hofbauer, J. and Sigmund, K., ``Evolutionary Games and
    Population Dynamics,'' (Cambridge Univ.\ Press, Cambridge, 1998).

\bibitem{lhn} Lieberman, E., Hauert, Ch., and Nowak, M.A., ``Evolutionary dynamics on graphs,''
             Nature {\bf 433}, 312 (2005).

\bibitem{tl} Lundh, T., ``Which ball is the roundest? - a suggested
tournament stability index,'', J. Quant. Anal. Sports {\bf 2}, No. 3, Article 
1 (2006).

\bibitem{hs1} Stern, H.S., ``On the Probability of Winning a Football Game,''
The American Statistician {\bf 45}, 179 (1991).

\bibitem{hs2} Stern, H.S., `Shooting Darts," in the column ``A Statistician
Reads the Sports Pages," Chance {\bf 10}, vol. 3, 16 (1997).

\bibitem{hs3} Stern, H.S., ``How Accurately Can Sports Outcomes Be
Predicted?," in the column ``A Statistician Reads the Sports Pages,"
Chance {\bf 10}, vol. 4, 19 (1997).

\bibitem{hs4} Stern, H.~S. and Mock, B.~R., ``College Basketball
Upsets: Will a 16-Seed Ever Beat a 1-Seed?," in the column ``A
Statistician Reads the Sports Pages," Chance {\bf 11}, No. 1, 26
(1998).


\bibitem{jw} Wesson, J., {\it The Science of Soccer}, (IOP, Bristol
  and Philadelphia, 2002).



\end{thebibliography}
\end{document}